\documentclass[prl,aps,
twocolumn,
superscriptaddress]{revtex4}

\usepackage{xcolor}
\usepackage{graphicx}

\usepackage[caption=false]{subfig}

\usepackage{amsmath}
\usepackage{booktabs}
\usepackage{rotating}
\usepackage{multirow}

\usepackage{bm}        
\usepackage{txfonts}   
\usepackage{hyperref}  

\date{\today}
\newcommand{\be}{\begin{eqnarray}}
\newcommand{\ee}{\end{eqnarray}}

\begin{document}
%

%
\title{$J/\psi$ production in proton-proton collisions at Spin Physics Detector energies of the JINR Nuclotron-based Ion Collider fAcility}

\author{Shubham Sharma}
\email[]{s.sharma.hep@gmail.com}
\affiliation{Moscow Institute of Physics and Technology, Dolgoprudny 141700, Russia}

\author{Alexey Aparin}
\email[]{aparin@jinr.ru}
\affiliation{Moscow Institute of Physics and Technology, Dolgoprudny $141700$, Russia}
\affiliation{Joint Institute for Nuclear Research, Dubna $141980$, Russia}
\affiliation{Institute for Nuclear Physics, Almaty, $050032$, Kazakhstan}

\date{\today}
%
\begin{abstract}
We investigate inclusive $J/\psi$ production in proton-proton collisions at tens of GeV $\sqrt{s}$ energy, relevant for forthcoming measurements with the Spin Physics Detector (SPD) at NICA. Simulations are performed using the PEGASUS event generator with transverse-momentum-dependent (TMD) gluon densities, comparing the recent KMR-based KL$'2025$ and CCFM-based LLM$'2024$ parametrizations. Differential cross sections in rapidity and transverse momentum exhibit smooth, stable behavior under renormalization-scale variation. Normalized $p_T$ spectra reveal distinct hardening patterns linked to the underlying gluon $k_T$ broadening in each model. The relative contributions of color-singlet and color-octet channels are also quantified, demonstrating the dominance of color-octet mechanisms in the SPD energy regime. These results provide the first detailed assessment of quarkonium production sensitivity to gluon TMDs near threshold, offering timely theoretical guidance for upcoming $J/\psi$ measurements at SPD/NICA.
%

\end{abstract}
%
\maketitle
%
%
{\it\textbf{1.  Motivation}}---
Quarkonium production in hadronic collisions serves as a powerful probe of Quantum Chromodynamics (QCD) across its perturbative and nonperturbative domains~\cite{ALICE:2021dtt,Arbuzov:2020cqg}. In particular, the production of the charmonium state $J/\psi$ is directly sensitive to gluon dynamics: at leading order it proceeds via gluon–gluon fusion, while its hadronization into a bound state involves the interplay between short-distance scattering and long-distance QCD effects~\cite{Arbuzov:2020cqg,physics5030044,Petrelli:1997ge}. 

Measurements of inclusive $J/\psi$ production over a broad range of center-of-mass energies have enabled stringent tests of theoretical frameworks such as the Color-Singlet Model (CSM)~\cite{Baier:1983va}, Nonrelativistic QCD (NRQCD) factorization~\cite{Bodwin:1994jh,Butenschoen:2012px}, and the Color-Evaporation Model (CEM)~\cite{Frawley:2008kk}. High-precision LHC measurements~\cite{ATLAS:2011aqv,LHCb:2021pyk,ALICE:2021dtt} have established a detailed picture of quarkonium production at multi-TeV energies, where global analyses combining color-singlet (CS) and color-octet (CO) mechanisms achieve good agreement with data.

At moderate collision energies, however, the situation is less constrained. Here, the relevant gluon momentum fractions are larger, the available phase space for high-$p_T$ recoils is reduced, and the sensitivity to the transverse motion of partons becomes clearer~\cite{Aidala:2012mv}. In this regime, the transverse-momentum-dependent (TMD) formalism provides a natural extension of QCD factorization, allowing direct access to the intrinsic $k_T$ structure and polarization of gluons inside the proton~\cite{physics5030044,Arbuzov:2020cqg}. Quarkonium production, particularly inclusive $J/\psi$, thus emerges as a promising observable to probe gluon TMDs and their evolution.

The forthcoming Spin Physics Detector (SPD) experiment at the NICA collider will explore proton–proton collisions at $\sqrt{s}\leq27~\mathrm{GeV}$ with polarized beams~\cite{Arbuzov:2020cqg,physics5030044}. This energy range bridges the gap between fixed-target and high-energy collider experiments, providing a uniquely clean environment to investigate gluon dynamics in the transition region where perturbative and nonperturbative effects overlap~\cite{SPDproto:2021hnm}.

In this work, we investigate inclusive $J/\psi$ production in proton–proton collisions for $\sqrt{s}\leq27~\mathrm{GeV}$ using the \textsc{PEGASUS} event generator, which implements TMD gluon densities in a $k_T$-factorized framework~\cite{Lipatov:2019oxs}. Two recent TMD parameterizations are employed: the KMR-based KL$'2025$~\cite{Kotikov:2025wft} and the CCFM-based LLM$'2024$~\cite{Lipatov:2024xni} sets. We present differential cross sections in transverse momentum and rapidity, including theoretical uncertainties from renormalization-scale variations. The normalized $p_T$ spectra reveal distinct behaviors for the two TMD sets, reflecting their different evolution patterns at SPD scale. Additionally, the relative contributions of CS and CO channels are analyzed across beam energies, elucidating how production mechanisms evolve from the SPD regime to the high-energy domain. These results establish the predictive foundations for quarkonium measurements at SPD/NICA and quantify the sensitivity of $J/\psi$ observables to the underlying gluon densities.
\newline\\
%
{\it\textbf{2.  Theoretical Framework}}---
Within the NRQCD factorization approach, 
the inclusive $J/\psi$ production cross section is written as a sum over intermediate $c\bar{c}$ states $n$ ~\cite{Bodwin:1994jh,Petrelli:1997ge},
\begin{equation}
d\sigma(pp \to J/\psi + X) = 
\sum_{n} d\hat{\sigma}(pp \to c\bar{c}[n] + X)\,
\langle \mathcal{O}^{J/\psi}[n] \rangle,
\end{equation}
where $d\hat{\sigma}$ represents the perturbatively calculable short-distance coefficient (SDC) for the production of a heavy-quark pair in state $[n]$, and 
$\langle \mathcal{O}^{J/\psi}[n] \rangle$ denotes the long-distance matrix element (LDME) governing its nonperturbative transition into a physical $J/\psi$ meson.

At leading order in $\alpha_s$, two classes of partonic subprocesses contribute:

\begin{enumerate}
\item[(i)] {CS mechanism:}
\begin{equation}
g^* + g^* \to c\bar{c}\big[{}^3S_1^{(1)}\big] + g,
\end{equation}
where the $c\bar{c}$ pair is produced directly in a CS configuration with the same quantum numbers as the $J/\psi$~\cite{Petrelli:1997ge}. The final-state gluon ensures color conservation and carries part of the recoil momentum.

\item[(ii)] {CO mechanisms:}
\begin{equation}
g^* + g^* \to c\bar{c}[n], \quad
n \in \big\{{}^1S_0^{(8)},\, {}^3S_1^{(8)},\, {}^3P_J^{(8)}\, (J=0,1,2)\big\},
\end{equation}
where the $c\bar{c}$ pair is produced in a color-octet state and subsequently evolves into the physical $J/\psi$ through soft-gluon emissions, with the $P$-wave octet states undergoing electric dipole (E1) transitions treated in the formalism of Ref.~\cite{Baranov:2016mka}.
%
%
These $2 \to 1$ subprocesses contribute at the same order in $\alpha_s$ as the CS $2 \to 2$ process.
\end{enumerate}

Both CS and CO channels are implemented in the 
\textsc{PEGASUS}~\cite{Lipatov:2019oxs} event generator within the TMD factorization framework, 
where the initial gluons are off-shell ($g^*$) and carry intrinsic transverse momentum $k_T$. 
The generated event samples include all relevant intermediate $c\bar{c}[n]$ states 
($[{}^3S_1^{(1)}]$, $[{}^1S_0^{(8)}]$, $[{}^3S_1^{(8)}]$, $[{}^3P_J^{(8)}]$), 
allowing a consistent treatment of both color and spin dynamics of $J/\psi$ production at SPD/NICA energies.

In this study, the LDMEs are set to unity to isolate the kinematic dependence of the SDCs and the impact of the gluon TMDs on the cross section. The partonic cross sections are convoluted with TMD gluon densities evolved via the CCFM and KMR schemes. 
The renormalization and factorization scales are chosen as $\mu^2 = m_T^2 = M_{J/\psi}^2 + p_T^2$, 
with the $J/\psi$ mass consistent with PDG values \cite{ParticleDataGroup:2018ovx,Lipatov:2019oxs}. 

Monte Carlo event generation and phase-space integration are performed using 
\textsc{PEGASUS}, which provides \texttt{.lhe} samples containing weighted events
according to the $k_T$-factorization formalism \cite{Lipatov:2019oxs, Alwall:2006yp}. 
Twelve statistically independent runs, each with $4\times10^5$ events, are produced to ensure stable predictions of 
differential distributions in $p_T$, $y$, and $\sqrt{s}$. For consistent comparison across different TMD gluon sets, the results are normalized to the total hadronic cross section where needed.

This framework provides a unified and quantitative basis to explore the sensitivity of inclusive $J/\psi$ production to the underlying gluon dynamics in the SPD/NICA energy range.
%




%
\begin{figure}[t]
  \centering
  \subfloat[$\sqrt{s}=9~\mathrm{GeV}$]{
    \includegraphics[width=\columnwidth]{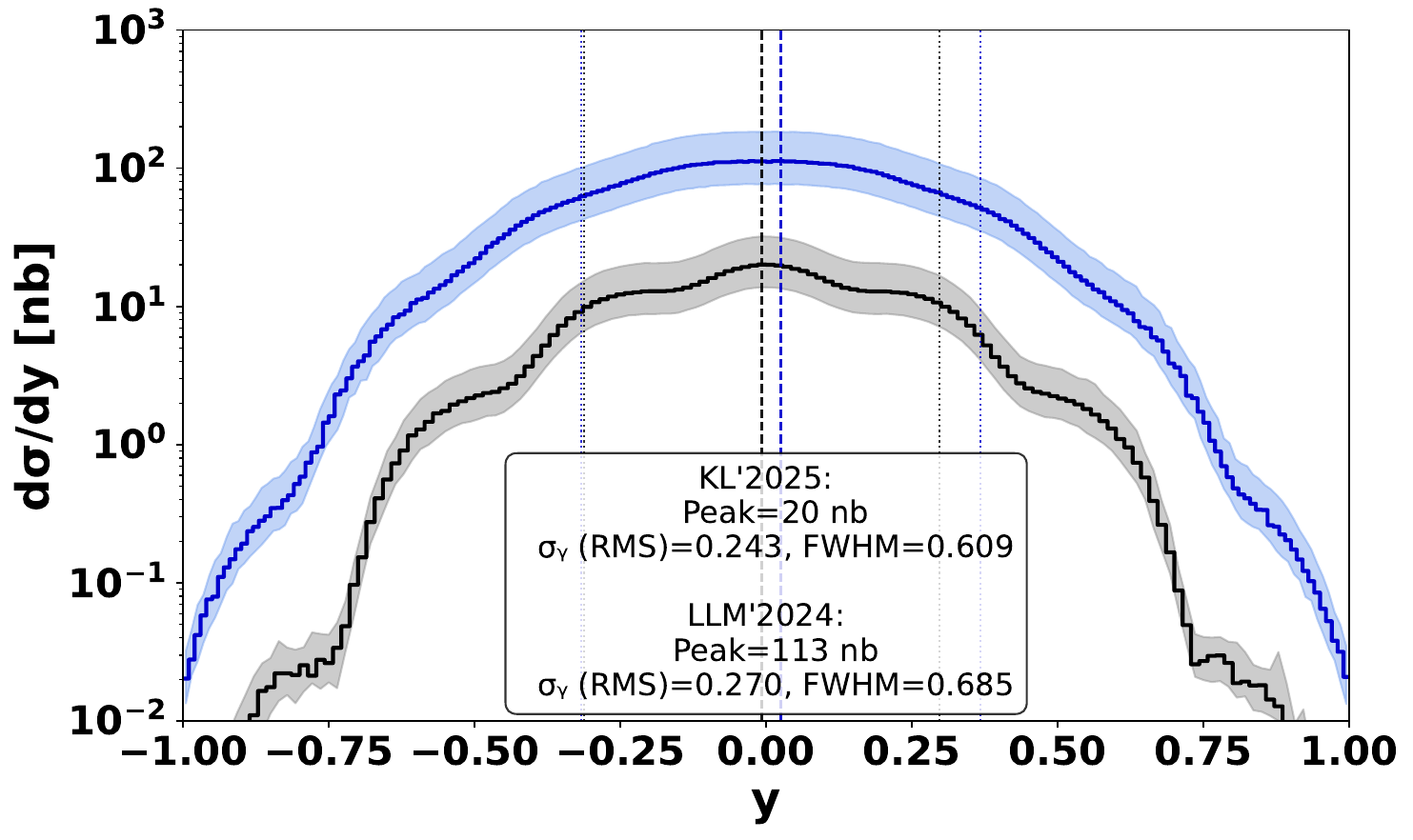}}

  \subfloat[$\sqrt{s}=18~\mathrm{GeV}$]{
    \includegraphics[width=\columnwidth]{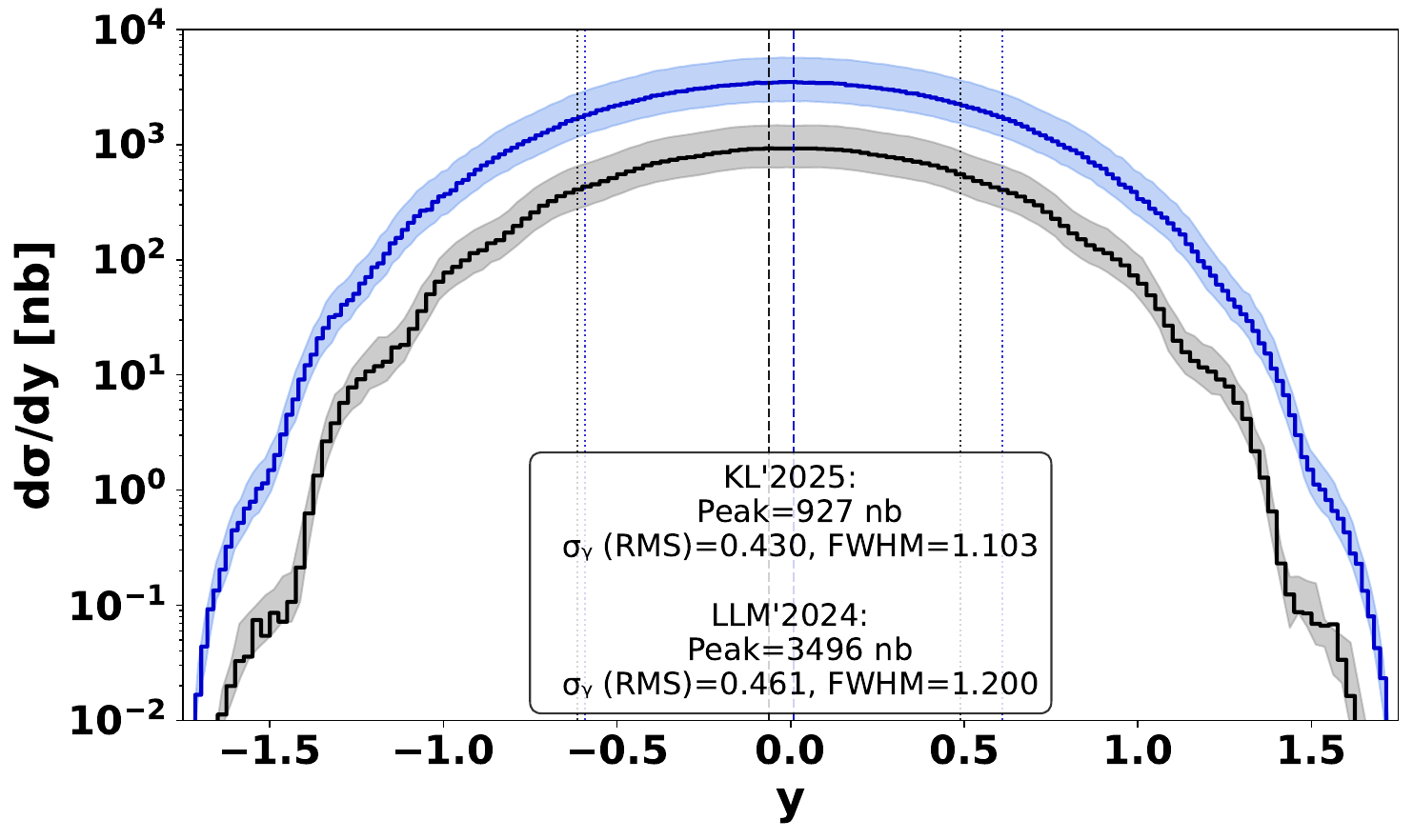}
  }

  \subfloat[$\sqrt{s}=27~\mathrm{GeV}$]{
    \includegraphics[width=\columnwidth]{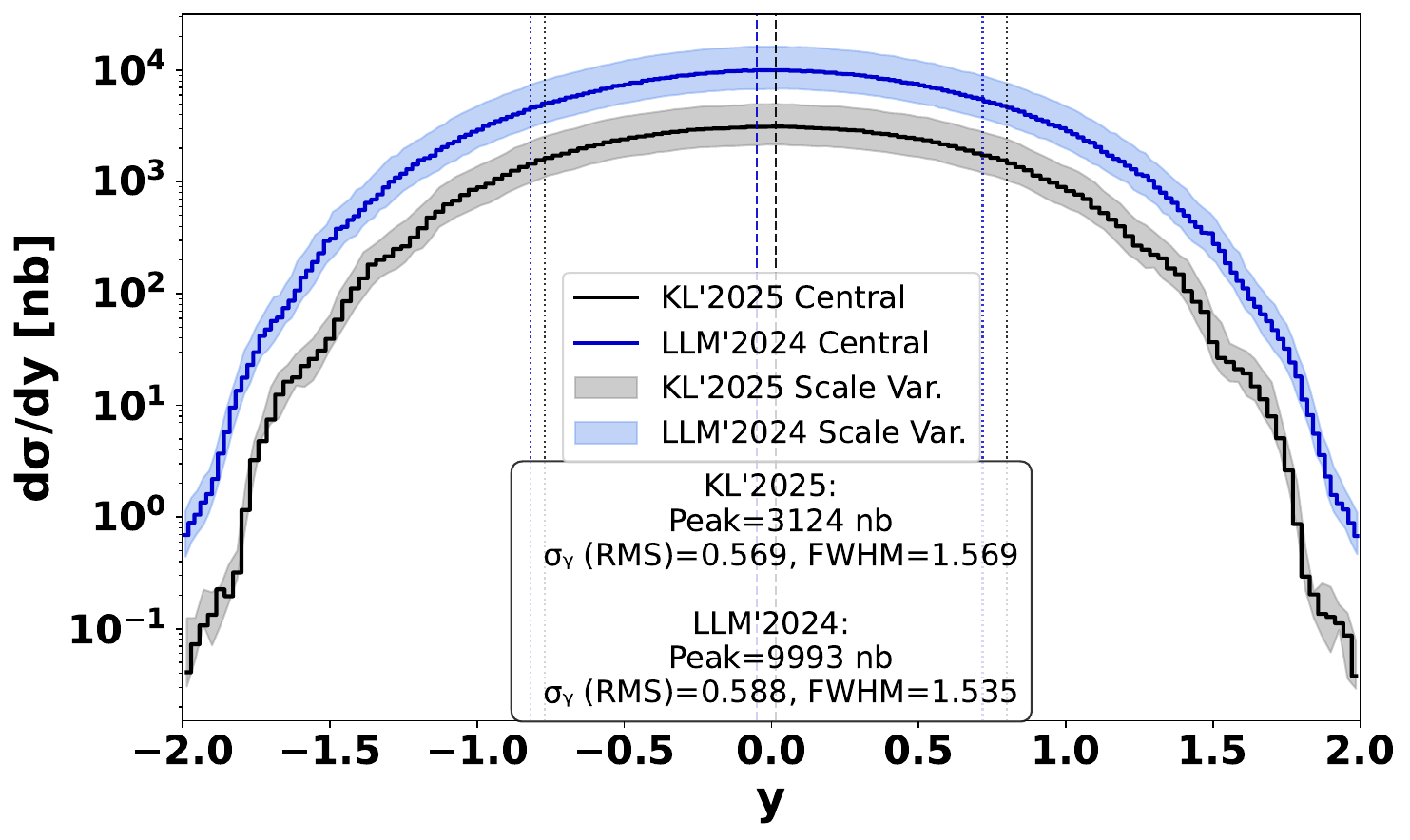}
  }
  \caption{
  Rapidity distributions ${d\sigma}/{dy}$ for inclusive $J/\psi$ production in pp collisions obtained using the KL$'2025$~\cite{Kotikov:2025wft} and LLM$'2024$~\cite{Lipatov:2024xni} gluon densities:
  (a) $\sqrt{s}=9~\mathrm{GeV}$,
  (b) $\sqrt{s}=18~\mathrm{GeV}$,
  (c) $\sqrt{s}=27~\mathrm{GeV}$.
The shaded bands represent the uncertainty from renormalization-scale variation. A clear broadening and enhancement of the distributions with increasing $\sqrt{s}$ reflect the expanding rapidity phase space and growing gluon-gluon luminosity at higher energies.}
  \label{fig:dsigmadY}
\end{figure}

%
\begin{figure}[t]
  \centering
\subfloat[$\sqrt{s}=9~\mathrm{GeV}$]{
    \includegraphics[width=\columnwidth]{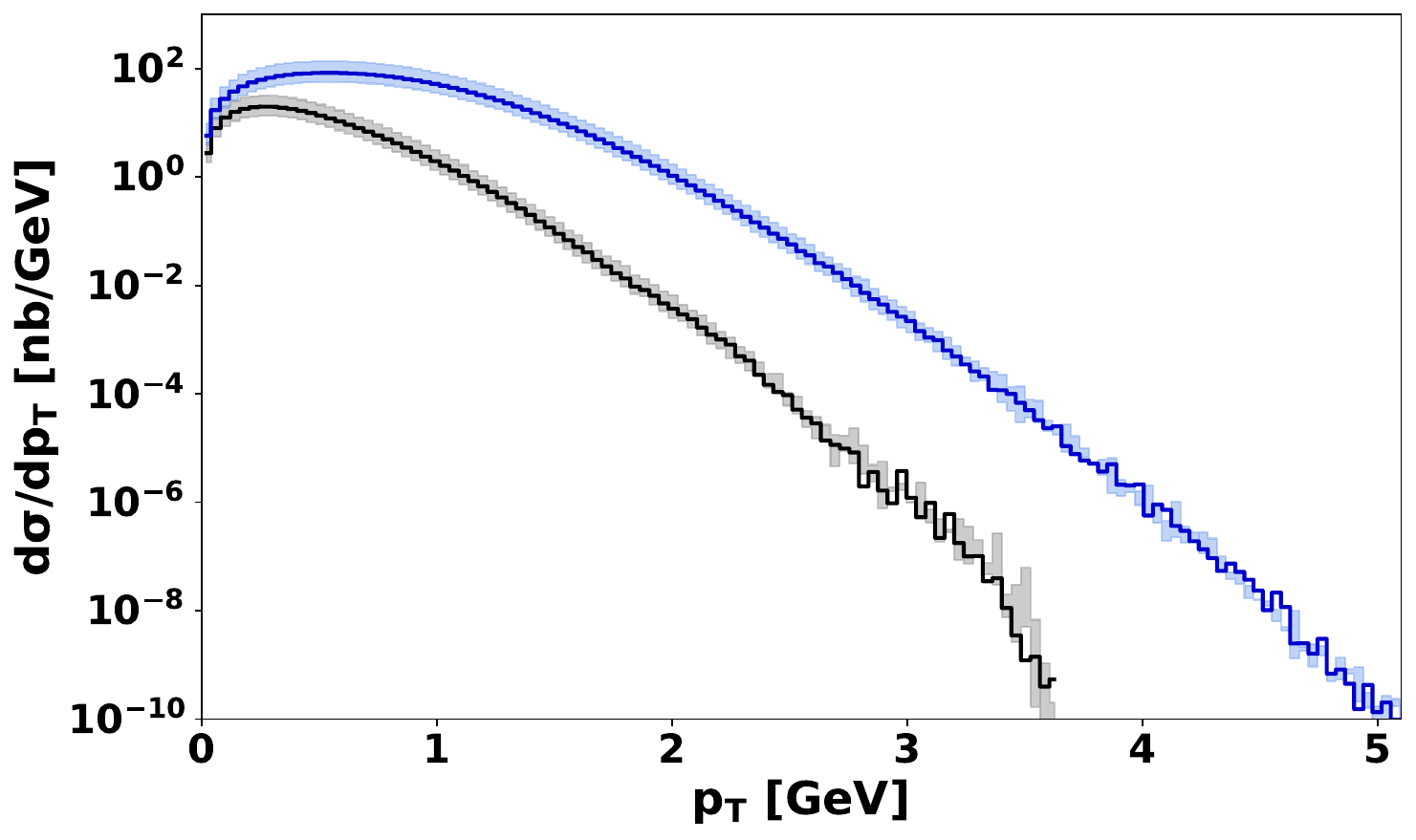}
  }
  
\subfloat[$\sqrt{s}=18~\mathrm{GeV}$]{
    \includegraphics[width=\columnwidth]{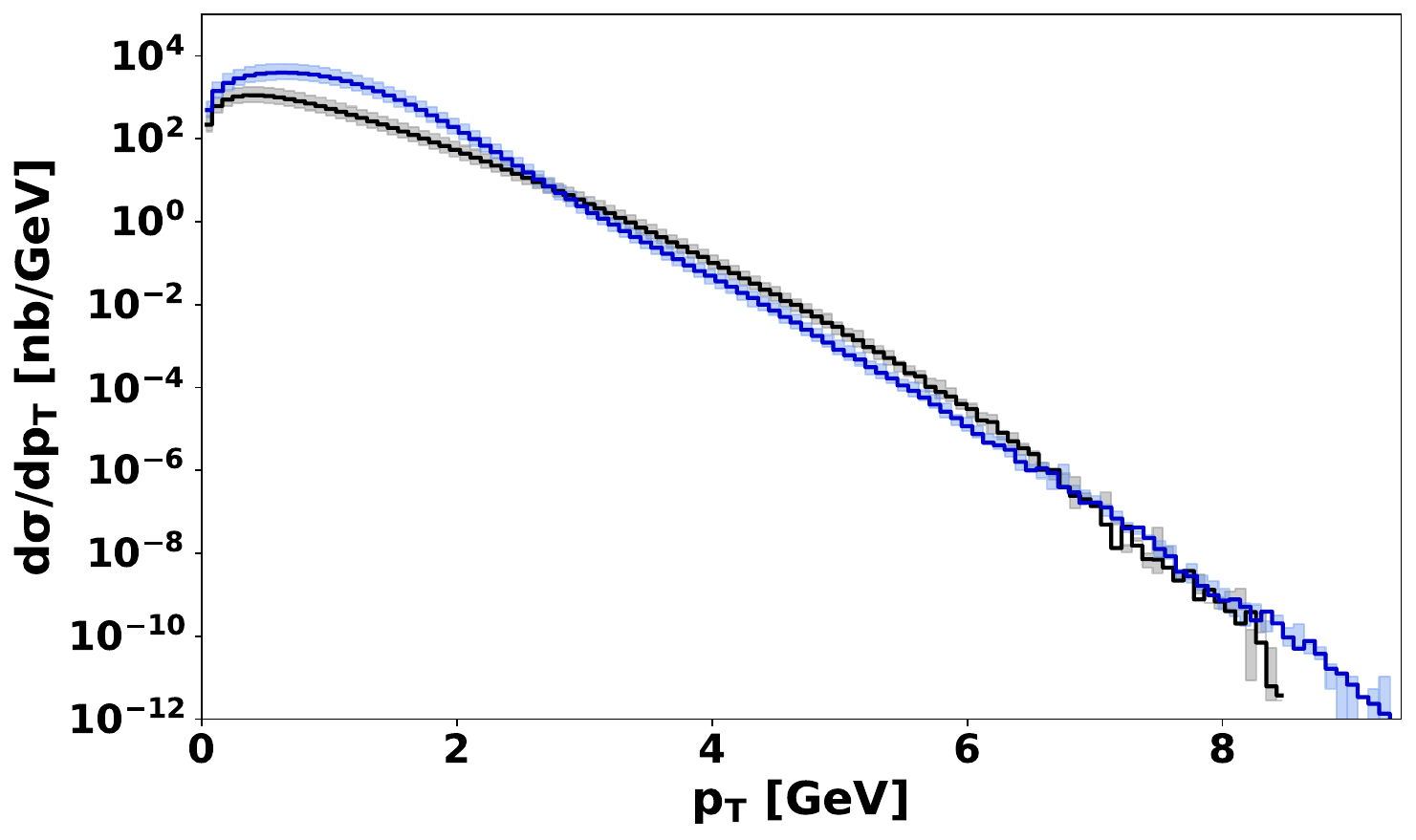}
  }
  
\subfloat[$\sqrt{s}=27~\mathrm{GeV}$]{
    \includegraphics[width=\columnwidth]{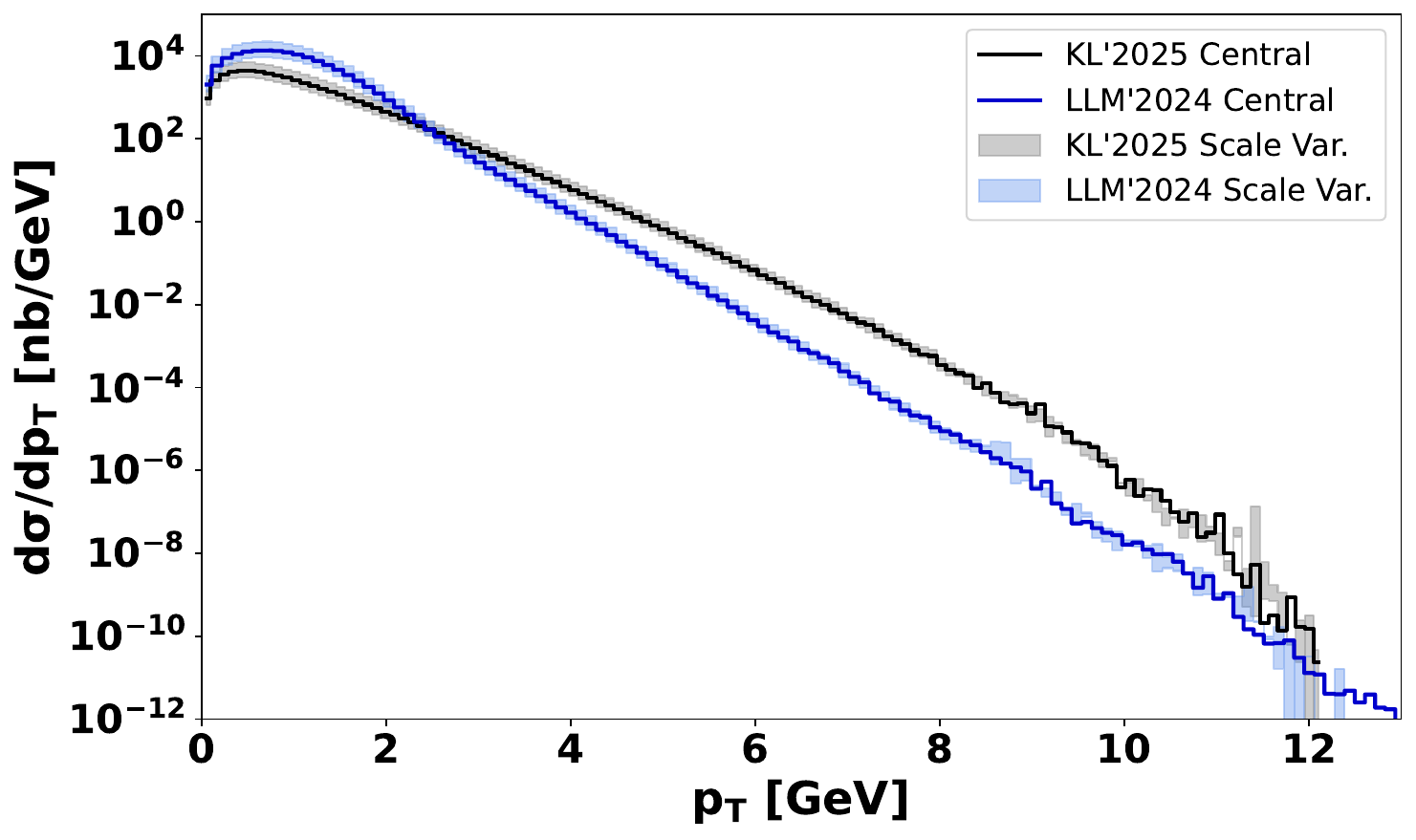}
  }
  \caption{Differential cross sections ${d\sigma}/{dp_T}$ for inclusive $J/\psi$ production in pp collisions obtained using the KL$'2025$~\cite{Kotikov:2025wft} and LLM$'2024$~\cite{Lipatov:2024xni} gluon densities:
  (a) $\sqrt{s}=9~\mathrm{GeV}$,
  (b) $\sqrt{s}=18~\mathrm{GeV}$, and
  (c) $\sqrt{s}=27~\mathrm{GeV}$.
  The shaded bands indicate the uncertainty due to renormalization-scale variation. The comparison illustrates the evolution of the $p_T$ spectrum with increasing collision energy.}
  \label{fig:dsig_dpt}
\end{figure}
{\it\textbf{3.  Results and Discussion}}---
Figure~\ref{fig:dsigmadY} shows the rapidity distributions $d\sigma/dy$ for inclusive $J/\psi$ production at $\sqrt{s}=9$, $18$, and $27~\mathrm{GeV}$. The spectra exhibit a symmetric and centrally peaked structure across all studied energies, with the peak position remaining close to mid-rapidity ($|y_{\text{peak}}|<0.07$), consistent with the expected symmetry of proton--proton collisions in this regime. 

A systematic broadening of the rapidity distribution with increasing energy is observed for both KL$'2025$ and LLM$'2024$ sets, as reflected by the rising RMS and FWHM values. The width roughly doubles from $\sqrt{s}=9$ to $27~\text{GeV}$, indicating the progressive opening of the rapidity phase space and the expanding kinematic reach in gluon momentum fractions. Concurrently, the overall magnitude of the distribution increases, reflecting the enhanced gluon–gluon luminosity that drives quarkonium production from the near-threshold to the intermediate-energy regime accessible at SPD. From the underlying event kinematics, we find that at $\sqrt{s}=27$ GeV the dominant contribution arises from a moderate-$x$ region, with median values $x\approx0.13$–0.16 and more than $90\%$ of events satisfying $x\leq0.3$, while only a small fraction probes the very low-$x$ domain.

The comparison between the two gluon density parameterizations reveals that LLM$'2024$ systematically gives higher cross sections at all rapidity values, while maintaining a slightly broader shape than KL$'2025$. This difference originates from the broader intrinsic transverse-momentum width and the softer small-$x$ gluon behavior encoded in the LLM$'2024$ evolution, which enhance central production rates. At the lowest energy, $\sqrt{s}=9~\mathrm{GeV}$, both models show a mild suppression around $y\approx0$, indicative of near-threshold kinematic constraints on the accessible gluon flux.

The shaded bands represent the theoretical uncertainty estimated by varying the renormalization scale around its central value, $\mu_R=m_T$, by a factor of two, i.e., $\mu_R = m_T/2$ and $\mu_R = 2m_T$, while keeping the factorization scale fixed at $\mu_F = m_T$. The quantitative spread and overall behavior of these bands remain stable across the considered energy range. It is worth noting that in the CCFM-based \(k_T\)-factorization approach, the factorization scale is intrinsically tied to the evolution variable of the unintegrated gluon density and therefore should not be treated as an independent parameter~\cite{Lipatov:2024xni}. Consequently, variations of the factorization scale do not provide a meaningful estimate of theoretical uncertainty and are not considered in the present analysis.

The two gluon-density sets thus yield nearly identical rapidity shapes but notably different normalizations, indicating that the overall production rate is primarily controlled by the gluon-luminosity normalization, for a fixed choice of non-perturbative input, whereas the shape is determined by the kinematic mapping of the partonic subprocess. This systematic comparison highlights that, within the SPD energy domain, the $y$-distribution serves as a sensitive probe of both the gluon-density normalization and the evolution pattern encoded in different unintegrated PDF parameterizations. The observed energy dependence of the width and normalization provides a consistent phenomenological signature of gluon-gluon fusion dominance and establishes a quantitative baseline for forthcoming SPD measurements of quarkonium production.

It is interesting to note that, at comparable energies, exclusive $J/\psi$ production in proton-proton ultraperipheral collisions yields $d\sigma/dy$ values that are $10^{-5}$-$10^{-4}$ times smaller than those reported here~\cite{Goloskokov:2025hsk}. The large difference originates from the dominance of electromagnetic interactions in ultraperipheral processes. Nevertheless, such measurements provide a valuable complementary perspective on $J/\psi$ production dynamics across different interaction regimes.


The transverse-momentum spectra, \(d\sigma/dp_T\), presented in Fig.~\ref{fig:dsig_dpt}, exhibit the expected steep falloff with increasing \(p_T\). Both gluon-density models reproduce this trend, but the spectral shapes reveal distinct features. At $\sqrt{s}=9~\mathrm{GeV}$, the spectrum is limited to low $p_T$, reflecting the restricted phase space near threshold. With \(p_T\), KL$'2025$ prediction shows a relatively smaller initial amplitude and harder tail near the end. As the energy increases, the spectra broaden from about $p_T\!\leq\!5~\mathrm{GeV}$ at $\sqrt{s}=9~\mathrm{GeV}$ to nearly $p_T\!\approx\!12~\mathrm{GeV}$ at $\sqrt{s}=27~\mathrm{GeV}$, illustrating the gradual transition from a soft, near-threshold domain to a more perturbative regime. Growth in overall normalization is consistent with the increasing gluon luminosity and available phase space. At these energies, LLM$'2024$ prediction relatively yields the initial higher amplitude and harder tail with \(p_T\) compared to the KL$'2025$ and a crossover region around $2~\mathrm{GeV} < p_T < 3~\mathrm{GeV}$ is observed, signaling the onset of enhanced partonic activity beyond which KL$'2025$ dominates. These features confirm that the \(p_T\) spectrum at SPD energies is sensitive to the magnitude and shape of gluon TMDs in the moderate-\(x\) domain (\(x \approx 10^{-2}\!-\!10^{-1}\)).

It is important to mention that LLM'24 density show predicitons beyond the maximum $p_T$ range show here but fluctuations in its precision increases rapidly for ${d\sigma}/{dp_T}$ values below $10^{-12}~\mathrm{nb/GeV}$.

\begin{figure}
    \centering
\includegraphics[width=\columnwidth]{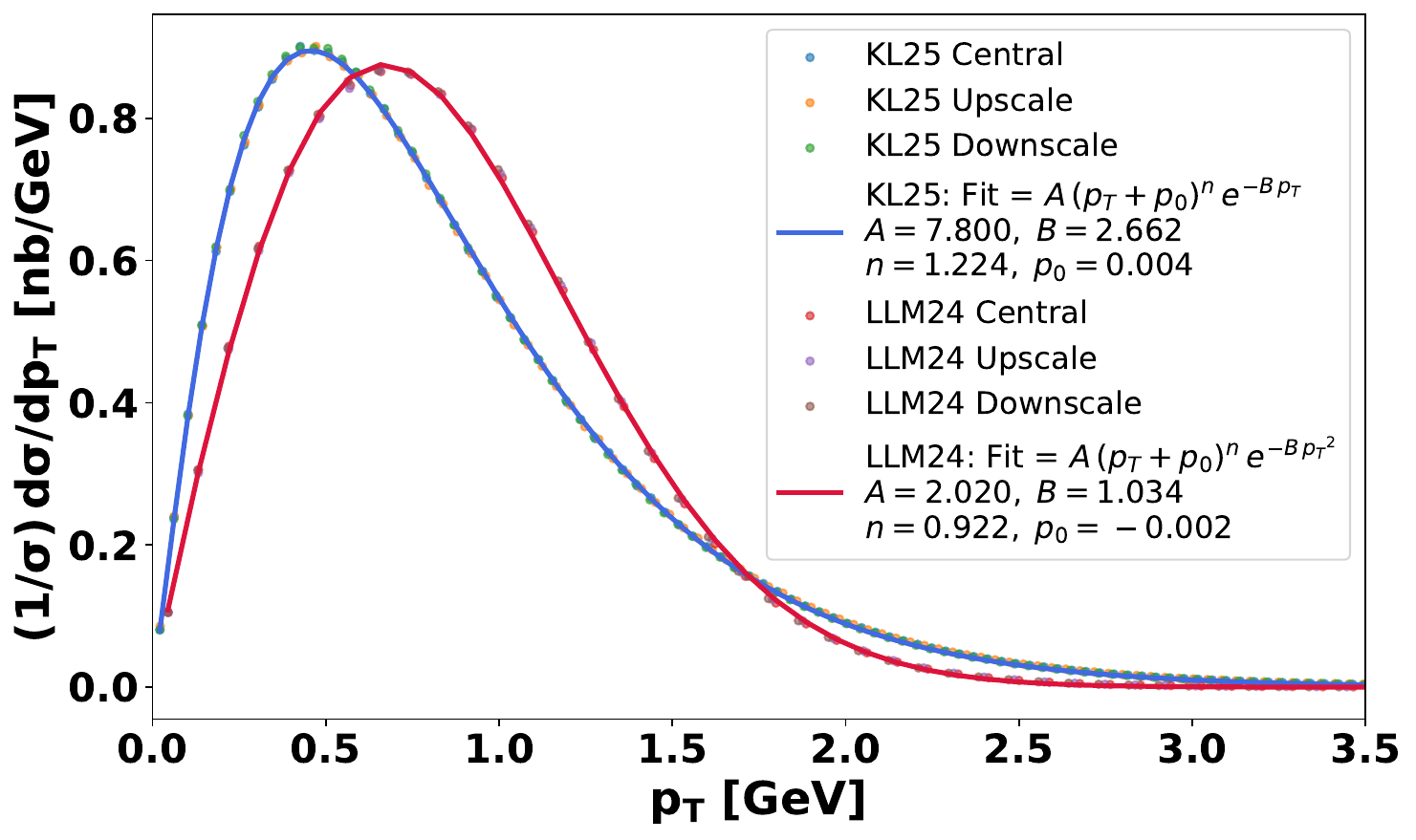}
\caption{Normalized differential cross sections $(1/\sigma)\, d\sigma/dp_T$ for $J/\psi$ production in $pp$ collisions at $\sqrt{s}=27~\mathrm{GeV}$ obtained using the KL$'2025$~\cite{Kotikov:2025wft} and LLM$'2024$~\cite{Lipatov:2024xni} gluon densities. The solid curves represent empirical fits to the simulated spectra with parameters shown in the legend. The distributions are normalized to emphasize differences in spectral shape between the two TMD sets.
}
\label{fig:ptnorm}
\end{figure}

To further quantify the model differences, normalized $J/\psi$ transverse-momentum spectra are compared in Fig.~\ref{fig:ptnorm}. The data are fitted with empirical functions of the form 
\(
A(p_T + p_0)^n e^{-B p_T}
\)
for KL$'2025$ and 
\(
A(p_T + p_0)^n e^{-B p_T^2}
\)
for LLM$'2024$, yielding excellent agreement across the full $p_T$ range. 

The KL$'2025$ fit, characterized by a larger parameter $n$, corresponds to a slightly harder rise of the spectrum, whereas the LLM$'2024$ form exhibits a smoother yet faster falloff at high $p_T$ due to its Gaussian-like dependence, even though the associated parameter $B$ is smaller. These distinct behaviors directly reflect the different gluon $k_T$-broadening mechanisms intrinsic to each TMD parametrization.

%

\begin{figure}
    \centering
\includegraphics[width=\columnwidth]{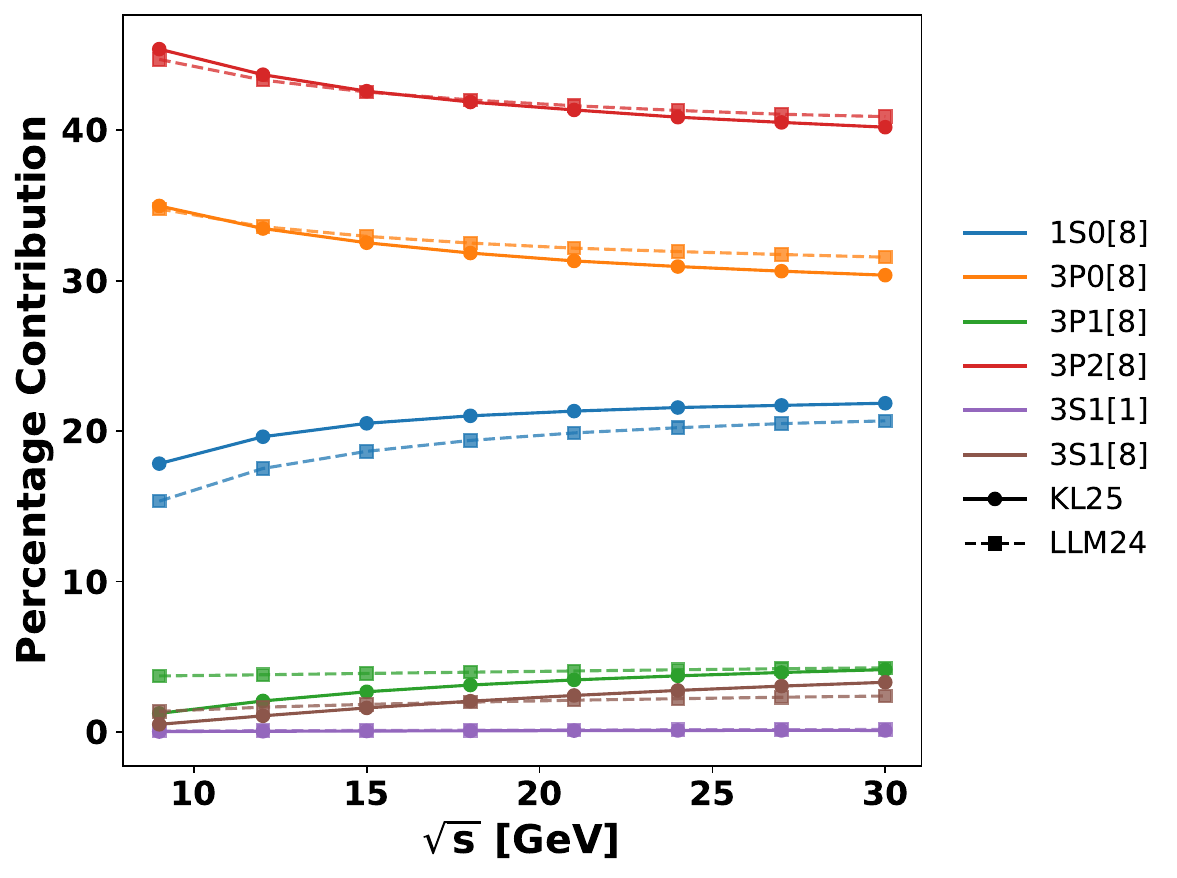}
\caption{Relative contributions of different intermediate $c\bar{c}$ states to the total $J/\psi$ production cross section in $pp$ collisions as a function of $\sqrt{s}$ for the KL$'2025$~\cite{Kotikov:2025wft} and LLM$'2024$~\cite{Lipatov:2024xni} gluon densities. The dominance pattern $^3P_{2}^{[8]}$$>$$^3P_{0}^{[8]}$$>$$^1S_{0}^{[8]}$$>$$^3P_{1}^{[8]}$$>$$^3S_{1}^{[8]}$$>$$^3S_{1}^{[1]}$ persists across the entire SPD energy range, with minimal model dependence.}
\label{fig:statefrac}
\end{figure}

Fig.~\ref{fig:statefrac} shows the relative contributions of the various \(c\bar{c}\) intermediate states to the total \(J/\psi\) yield. Both TMD models predict nearly identical channel hierarchies, with CO states-particularly \(^3P_{2}^{[8]}\), \(^3P_{0}^{[8]}\), and \(^1S_{0}^{[8]}\)-dominating across the entire SPD energy range. The CS \(^3S_1^{[1]}\) contribution remains below the percent level even at \(\sqrt{s}=30~\mathrm{GeV}\). This confirms the essential role of CO mechanisms in \(J/\psi\) production at intermediate energies.
 
\begin{figure}
    \centering
\includegraphics[width=\columnwidth]{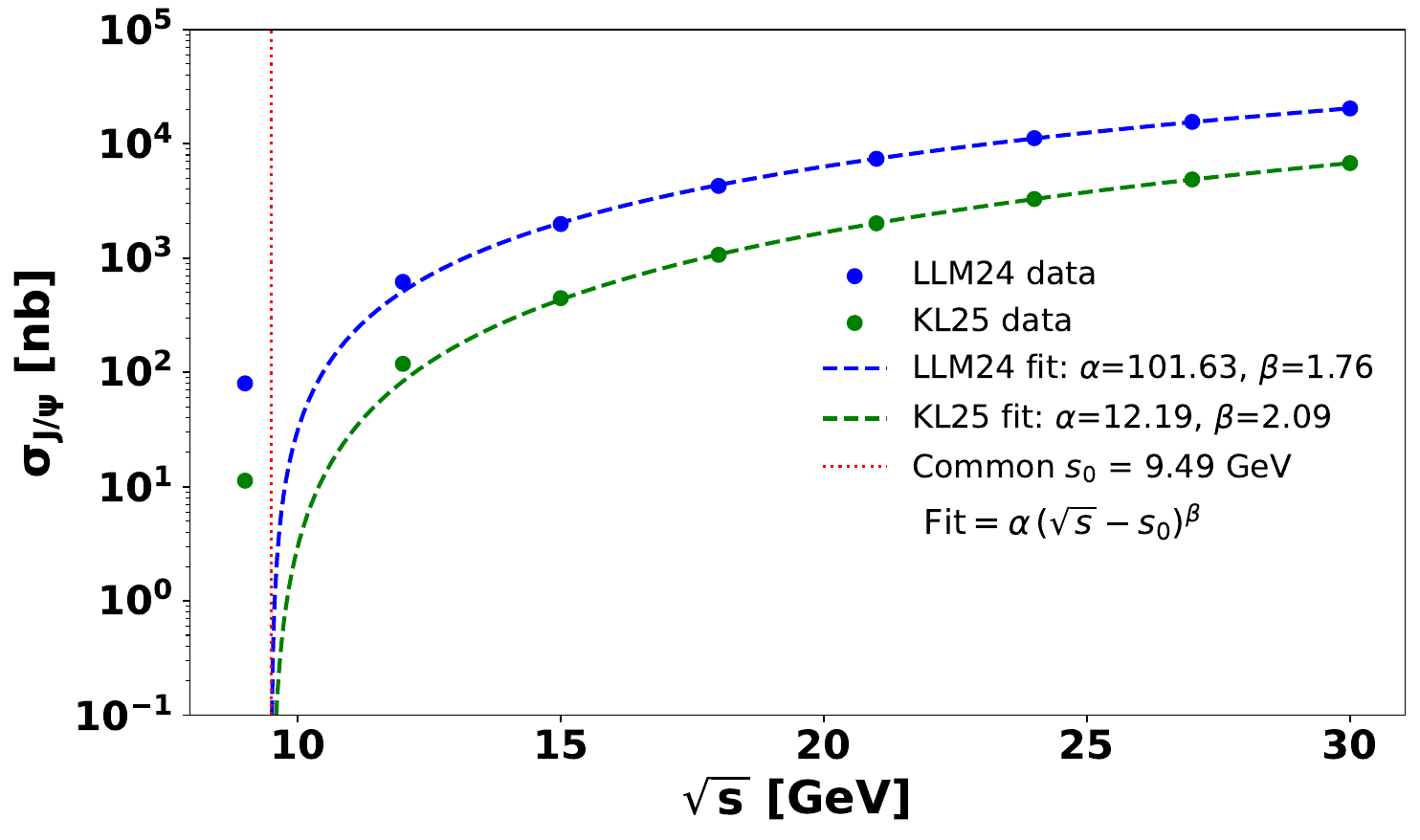}
\caption{Total inclusive $J/\psi$ production cross section in $pp$ collisions as a function of the center-of-mass energy $\sqrt{s}$, obtained with the KL$'2025$~\cite{Kotikov:2025wft} and LLM$'2024$~\cite{Lipatov:2024xni} gluon densities. The cross section increases with $\sqrt{s}$, reflecting the growth of gluon luminosity. The shifted power-law fit provides an excellent description for $\sqrt{s}\geq12~\mathrm{GeV}$; the deviation at $\sqrt{s}=9~\mathrm{GeV}$ likely signals threshold or model-limit effects.}
\label{fig:sigma_vs_sqrtS}
\end{figure}

Finally, the total inclusive cross section as a function of \(\sqrt{s}\) is presented in Fig.~\ref{fig:sigma_vs_sqrtS}. The cross section rises monotonically with increasing energy and is well described by a shifted power-law form, $(\sigma_{J/\psi} = \alpha(\sqrt{s}-s_0)^{\beta})$, for \(\sqrt{s} \geq 12~\mathrm{GeV}\). The point at \(\sqrt{s}=9~\mathrm{GeV}\) lies below the extrapolated curve, suggesting possible near-threshold suppression effects. The LLM$'2024$ parametrization yields a higher normalization and a softer energy dependence (\(\beta \approx 1.8\)) compared with KL$'2025$ (\(\beta \approx 2.1\)), implying a slower rise of gluon-driven production with energy. Between \(12\) and \(30~\mathrm{GeV}\), the LLM$'2024$-to-KL$'2025$ scaling ratio decreases from about 2.5 to 2.0, highlighting the diminishing relative contribution from the LLM$'2024$ gluon density as \(\sqrt{s}\) increases. These results emphasize the strong sensitivity of \(J/\psi\) production to the underlying gluon TMDs in the near-threshold region accessible at SPD/NICA.


\par

%
%
%

%
{\it\textbf{4. Summary}}---
This work presents the first quantitative analysis of inclusive $J/\psi$ production in the SPD/NICA energy regime using modern TMD gluon densities. We have investigated the sensitivity of differential observables to scale variations, intermediate-state contributions, and the collision-energy dependence within both CCFM and KMR-based frameworks. The results establish a robust theoretical baseline for forthcoming SPD experimet and emphasize the need for refined gluon-density extractions in the low to intermediate energy region, where quarkonium production remains a uniquely sensitive probe of nonperturbative gluon dynamics.\\



{\it\textbf{Acknowledgements}}---
S.S. expresses sincere gratitude to Prof. A.V.~Lipatov for valuable clarifications regarding the use of TMD parton densities in \textsc{PEGASUS}, and to S.~Puhan for helpful discussions. This work was done with support from Ministry of Science and Higher Education of Russian Federation under State Assignment № 075-03-2025-662 from 17.01.2025
and by the Committee of Science of the Ministry of Science and Higher Education of the Republic of Kazakhstan (Grant No. BR21881941).




\bibliography{ref_SS.bib}

@article{ALICE:2021dtt,
    author = "Acharya, Shreyasi and others",
    collaboration = "ALICE",
    title = "{Inclusive $\text {J}/\psi $ production at midrapidity in pp collisions at $\sqrt{s} = 13$ TeV}",
    eprint = "2108.01906",
    archivePrefix = "arXiv",
    primaryClass = "nucl-ex",
    reportNumber = "CERN-EP-2021-160",
    doi = "10.1140/epjc/s10052-021-09873-4",
    journal = "Eur. Phys. J. C",
    volume = "81",
    number = "12",
    pages = "1121",
    year = "2021"
}

@article{Arbuzov:2020cqg,
    author = "Arbuzov, A. and others",
    title = "{On the physics potential to study the gluon content of proton and deuteron at NICA SPD}",
    eprint = "2011.15005",
    archivePrefix = "arXiv",
    primaryClass = "hep-ex",
    doi = "10.1016/j.ppnp.2021.103858",
    journal = "Prog. Part. Nucl. Phys.",
    volume = "119",
    pages = "103858",
    year = "2021"
}

@article{Baier:1983va,
    author = "Baier, R. and Ruckl, R.",
    title = "{Hadronic Collisions: A Quarkonium Factory}",
    reportNumber = "BI-TP 83/02",
    doi = "10.1007/BF01572254",
    journal = "Z. Phys. C",
    volume = "19",
    pages = "251",
    year = "1983"
}

@article{Bodwin:1994jh,
    author = "Bodwin, Geoffrey T. and Braaten, Eric and Lepage, G. Peter",
    title = "{Rigorous QCD analysis of inclusive annihilation and production of heavy quarkonium}",
    eprint = "hep-ph/9407339",
    archivePrefix = "arXiv",
    reportNumber = "ANL-HEP-PR-94-24, FERMILAB-PUB-94-073-T, NUHEP-TH-94-5",
    doi = "10.1103/PhysRevD.55.5853",
    journal = "Phys. Rev. D",
    volume = "51",
    pages = "1125--1171",
    year = "1995",
    note = "[Erratum: Phys.Rev.D 55, 5853 (1997)]"
}

@article{Butenschoen:2012px,
    author = "Butenschoen, Mathias and Kniehl, Bernd A.",
    title = "{J/psi polarization at Tevatron and LHC: Nonrelativistic-QCD factorization at the crossroads}",
    eprint = "1201.1872",
    archivePrefix = "arXiv",
    primaryClass = "hep-ph",
    reportNumber = "DESY-12-001",
    doi = "10.1103/PhysRevLett.108.172002",
    journal = "Phys. Rev. Lett.",
    volume = "108",
    pages = "172002",
    year = "2012"
}

@article{Frawley:2008kk,
    author = "Frawley, Anthony D. and Ullrich, T. and Vogt, R.",
    title = "{Heavy flavor in heavy-ion collisions at RHIC and RHIC II}",
    eprint = "0806.1013",
    archivePrefix = "arXiv",
    primaryClass = "nucl-ex",
    reportNumber = "LLNL-JRNL-402664, BNL-80239-2008-JA",
    doi = "10.1016/j.physrep.2008.04.002",
    journal = "Phys. Rept.",
    volume = "462",
    pages = "125--175",
    year = "2008"
}

@article{ATLAS:2011aqv,
    author = "Aad, Georges and others",
    collaboration = "ATLAS",
    title = "{Measurement of the differential cross-sections of inclusive, prompt and non-prompt $J/\psi$ production in proton-proton collisions at $\sqrt{s}=7$ TeV}",
    eprint = "1104.3038",
    archivePrefix = "arXiv",
    primaryClass = "hep-ex",
    reportNumber = "CERN-PH-EP-2011-041",
    doi = "10.1016/j.nuclphysb.2011.05.015",
    journal = "Nucl. Phys. B",
    volume = "850",
    pages = "387--444",
    year = "2011"
}

@article{LHCb:2021pyk,
    author = "Aaij, R. and others",
    collaboration = "LHCb",
    title = "{Measurement of $J/\psi$ production cross-sections in $pp$ collisions at $\sqrt{s}=5$ TeV}",
    eprint = "2109.00220",
    archivePrefix = "arXiv",
    primaryClass = "hep-ex",
    reportNumber = "LHCb-PAPER-2021-020, CERN-EP-2021-156",
    doi = "10.1007/JHEP11(2021)181",
    journal = "JHEP",
    volume = "11",
    pages = "181",
    year = "2021"
}

@article{Kotikov:2025wft,
    author = "Kotikov, A. V. and Lipatov, A. V.",
    title = "{Updating TMD parton densities in a proton within the Kimber-Martin-Ryskin approach}",
    eprint = "2502.19126",
    archivePrefix = "arXiv",
    primaryClass = "hep-ph",
    doi = "10.1103/PhysRevD.111.094009",
    journal = "Phys. Rev. D",
    volume = "111",
    number = "9",
    pages = "094009",
    year = "2025"
}

@article{Lipatov:2024xni,
    author = "Lipatov, A. V. and Lykasov, G. I. and Malyshev, M. A.",
    title = "{Refined TMD Gluon Density in a Proton from the HERA and LHC Data}",
    eprint = "2404.09550",
    archivePrefix = "arXiv",
    primaryClass = "hep-ph",
    doi = "10.1134/S0021364024601234",
    journal = "JETP Lett.",
    volume = "119",
    number = "11",
    pages = "828--833",
    year = "2024"
}

@inproceedings{Goloskokov:2025hsk,
    author = "Goloskokov, S. V. and Xie, Ya-Ping",
    title = "{Exclusive $J/?$ Production and Gluon GPDs Effects}",
    eprint = "2510.14572",
    archivePrefix = "arXiv",
    primaryClass = "hep-ph",
    month = "10",
    year = "2025"
}

@article{Lipatov:2019oxs,
    author = "Lipatov, A. V. and Malyshev, M. A. and Baranov, S. P.",
    title = "{Particle Event Generator: A Simple-in-Use System PEGASUS version 1.0}",
    eprint = "1912.04204",
    archivePrefix = "arXiv",
    primaryClass = "hep-ph",
    doi = "10.1140/epjc/s10052-020-7898-6",
    journal = "Eur. Phys. J. C",
    volume = "80",
    number = "4",
    pages = "330",
    year = "2020"
}

@article{Petrelli:1997ge,
    author = "Petrelli, Andrea and Cacciari, Matteo and Greco, Mario and Maltoni, Fabio and Mangano, Michelangelo L.",
    title = "{NLO production and decay of quarkonium}",
    eprint = "hep-ph/9707223",
    archivePrefix = "arXiv",
    reportNumber = "CERN-TH-97-142, DESY-97-090",
    doi = "10.1016/S0550-3213(97)00801-8",
    journal = "Nucl. Phys. B",
    volume = "514",
    pages = "245--309",
    year = "1998"
}

@article{Alwall:2006yp,
    author = "Alwall, J. and others",
    title = "{A Standard format for Les Houches event files}",
    eprint = "hep-ph/0609017",
    archivePrefix = "arXiv",
    reportNumber = "FERMILAB-PUB-06-337-T, CERN-LCGAPP-2006-03",
    doi = "10.1016/j.cpc.2006.11.010",
    journal = "Comput. Phys. Commun.",
    volume = "176",
    pages = "300--304",
    year = "2007"
}

@Article{physics5030044,
AUTHOR = {Guskov, Alexey and Datta, Amaresh and Karpishkov, Anton and Denisenko, Igor and Saleev, Vladimir},
TITLE = {Probing Gluons with the Future Spin Physics Detector},
JOURNAL = {Physics},
VOLUME = {5},
YEAR = {2023},
NUMBER = {3},
PAGES = {672--687},
URL = {https://www.mdpi.com/2624-8174/5/3/44},
ISSN = {2624-8174},
DOI = {10.3390/physics5030044}
}

@article{ParticleDataGroup:2018ovx,
    author = "Tanabashi, M. and others",
    collaboration = "Particle Data Group",
    title = "{Review of Particle Physics}",
    doi = "10.1103/PhysRevD.98.030001",
    journal = "Phys. Rev. D",
    volume = "98",
    number = "3",
    pages = "030001",
    year = "2018"
}

@article{Aidala:2012mv,
    author = "Aidala, Christine A. and Bass, Steven D. and Hasch, Delia and Mallot, Gerhard K.",
    title = "{The Spin Structure of the Nucleon}",
    eprint = "1209.2803",
    archivePrefix = "arXiv",
    primaryClass = "hep-ph",
    doi = "10.1103/RevModPhys.85.655",
    journal = "Rev. Mod. Phys.",
    volume = "85",
    pages = "655--691",
    year = "2013"
}

@article{Baranov:2016mka,
    author = "Baranov, S. P.",
    title = "{Possible solution of the quarkonium polarization problem}",
    reportNumber = "DESY-15-222",
    doi = "10.1103/PhysRevD.93.054037",
    journal = "Phys. Rev. D",
    volume = "93",
    number = "5",
    pages = "054037",
    year = "2016"
}

@article{SPDproto:2021hnm,
    author = "Abazov, V. M. and others",
    collaboration = "SPD proto",
    title = "{Conceptual design of the Spin Physics Detector}",
    eprint = "2102.00442",
    archivePrefix = "arXiv",
    primaryClass = "hep-ex",
    month = "1",
    year = "2021"
}

%
\end{document}